# Interplay between spin density wave and superconductivity in '122' iron pnictides: $^{57}$Fe Mössbauer study

A. Błachowski[1], K. Ruebenbauer[1*], J. Żukrowski[2], Z. Bukowski[3], M. Matusiak[3], and J. Karpinski[4]

[1] Mössbauer Spectroscopy Division, Institute of Physics, Pedagogical University
PL-30-084 Kraków, ul. Podchorążych 2, Poland

[2] AGH University of Science and Technology, Faculty of Physics and Applied Computer Science, Department of Solid State Physics
Ave. A. Mickiewicza 30, PL-30-059 Krakow, Poland

[3] Institute of Low Temperatures and Structure Research, Polish Academy of Sciences
PL-50-422 Wrocław, ul. Okólna 2, Poland

[4] Laboratory for Solid State Physics, ETH Zurich
CH-8093 Zurich, Switzerland

[*] Corresponding author: sfrueben@cyf-kr.edu.pl

**PACS:** 74.70.Xa, 75.30.Fv, 76.80.+y

**Keywords:** iron-based superconductors, spin density waves, Mössbauer spectroscopy

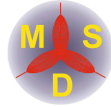

**Abstract**

Iron-based superconductors $Ba_{0.7}Rb_{0.3}Fe_2As_2$ and $CaFe_{1.92}Co_{0.08}As_2$ of the '122' family have been investigated by means of the 14.41-keV Mössbauer transition in $^{57}$Fe versus temperature ranging from the room temperature till 4.2 K. A comparison is made with the previously investigated parent compounds $BaFe_2As_2$ and $CaFe_2As_2$. It has been found that Mössbauer spectra of these superconductors are composed of the magnetically split component due to development of spin density wave (SDW) and non-magnetic component surviving even at lowest temperatures. The latter component is responsible for superconductivity. Hence, the superconductivity occurs in the part of the sample despite the sample is single phase. This phenomenon is caused by the slight variation of the dopant concentration across the sample (crystal).

## 1. Introduction

Parent compounds of the '122' family of the iron-based pnictide superconductors have formula $AFe_2As_2$ with (A=Ca, Ba, Sr, Eu) [1]. Parent compounds order magnetically by development of the spin density wave (SDW) generated by the iron atoms as far as the 3d magnetism is concerned. A development of SDW is accompanied by the slight distortion from the tetragonal to the orthorhombic unit cell. The longitudinal SDW propagates along the a-axis of the orthorhombic unit cell and it is incommensurate with the lattice period. Iron-arsenic layers are responsible for generation of SDW [2]. Superconductivity could be obtained applying either external pressure or doping (chemical pressure). Hole doping is achieved by substitution of A-atom by e.g. Rb [3]. Electron doping could be achieved by substitution of Fe by Co [4]. The iso-valence doping could be realized replacing arsenic by phosphorus [5]. SDW is strongly modified by doping, and it finally disappears at some level of dopant. Superconductivity appears in the regions without 3d magnetic order. Over-doping leads to disappearance of superconductivity and 3d magnetic order.

$Ba_{0.7}Rb_{0.3}Fe_2As_2$ and $CaFe_{1.92}Co_{0.08}As_2$ superconductors have been investigated versus temperature by means of the 14.41-keV Mössbauer transition in $^{57}Fe$. A comparison is made with the respective parents [6].

## 2. Experimental

The polycrystalline samples of $Ba_{0.7}Rb_{0.3}Fe_2As_2$ and $BaFe_2As_2$ were prepared by the solid state reaction at elevated temperature [3]. Single crystals of $CaFe_{1.92}Co_{0.08}As_2$ and $CaFe_2As_2$ compounds were grown by the tin flux method as described in Ref. [4]. Presence of the single phase was confirmed by powder X-ray diffraction method [3, 4]. Resisitivity has been measured by using the four-point method. Gold wires of 25 μm diameter were attached to the crystal using two components silver-based epoxy resin.

Mössbauer spectra were obtained on the powdered samples with the help of the $^{57}Co(Rh)$ source kept at room temperature. A vibration-free Janis Research SVT-400M cryostat was used to maintain absorber temperature with the accuracy better than 0.01 K. Spectra were collected by the MsAa-3 spectrometer calibrated by the laser operated Michelson interferometer. Spectra were processed by the *GMFPHARM* application of the Mosgraf-2009 suite within the transmission integral approximation [6, 7]. All spectral shifts are reported versus room temperature α-Fe.

## 3. Results

Figure 1 shows normalized (relative) resistivity plotted versus temperature for $CaFe_2As_2$ and $CaFe_{1.92}Co_{0.08}As_2$ compounds. Humps are due to the development of SDW (magnetic order). Transition to the superconducting state at 20 K is clearly seen for the $CaFe_{1.92}Co_{0.08}As_2$ compound. A transition to the superconducting state has been determined as 37 K for the $Ba_{0.7}Rb_{0.3}Fe_2As$ compound in Ref. [3]. The zero resistivity of the doped material is caused by the superconducting component making continuous conducting channels across the crystal (sample).

Figure 2a shows Mössbauer spectra for the parent compound $BaFe_2As_2$ versus temperature, while Figure 2b shows spectra of the superconductor $Ba_{0.7}Rb_{0.3}Fe_2As_2$ versus temperature as well. SDW appears at about 140 K and fully develops down to 130 K in the parent compound [6]. For material doped with Rb a dominant spectral component does not exhibit any hyperfine magnetic field even at lowest temperatures. A singlet characteristic of the parent compound at high temperature is replaced by doublet due to the electric quadrupole interaction on the iron nucleus induced by rubidium atoms replacing barium atoms. The isomer shift decreased by about 0.02 mm/s at room temperature indicating increased electron density on the iron nucleus with doping (see, Table and compare with Ref. [6]). However, about 15 % of the spectrum show some SDW at low temperature (superconducting region) indicating that some regions of the sample remain under-doped. There are two subsequent odd harmonics in SDW at 80 K and three close to magnetic saturation.

Spectra of $CaFe_2As_2$ are shown in Figure 3a, while the spectra of $CaFe_{1.92}Co_{0.08}As_2$ are shown in Figure 3b. SDW appears in the parent compound at 180 K and fully develops slightly above 80 K. The parent compound exhibits some quadrupole interaction on iron even in the high temperature tetragonal phase [6]. There are no significant changes of the isomer shift with doping. A spectral component with SDW amounts to 83 % at lowest temperatures for the doped material, while the non-magnetic (superconducting) component with the quadrupole interaction is the remainder. SDW contains six subsequent odd harmonics.

Figure 4 shows shape of SDW versus temperature for both parent compounds and for $CaFe_{1.92}Co_{0.08}As_2$. SDW for parents develops with lowering temperature from nearly isolated sheets with strong magnetization to almost rectangular shape. The maximum amplitudes of magnetization change a little with temperature. The rectangle is well developed for the Ba-based compound and somewhat less developed for the Ca-based compound at the lowest temperatures [6]. One can observe for the doped material that the maximum amplitude of SDW is not much different from the amplitude in the parent in the whole relevant temperature range. However, amplitudes of the high harmonics (9-th or 7-th depending on the temperature) are strongly enhanced leading to the rich structure of the SDW shape. Amplitudes of the subsequent harmonics do not diminish monotonously – in the absolute sense. Some evolution of the shape is observed below 80 K – mainly as filling of the weakly magnetized space.

Square roots from the mean squared amplitudes of SDW $\sqrt{\langle B^2 \rangle}$ are plotted versus temperature in Figure 5. A behavior of this quantity versus temperature has been discussed in detail for parents in Ref. [6]. One can observe that for the $CaFe_{1.92}Co_{0.08}As_2$ compound some upturn could be seen below 80 K due to the above mentioned filling of the weakly magnetized space by SDW at low temperature. The upturn occurs on the already flat part suggesting some additional electron spin coupling mechanism becoming effective below 80 K.

Substitution on the iron sites reduces significantly $\sqrt{\langle B^2 \rangle}$, while substitution on the A-site (Ba-site) has very weak influence on this parameter, but strong effect on the amount of the magnetic (SDW) component.

4. Conclusions

Coexistence of the magnetic order (3d electron generated SDW) and superconductivity could occur within the same sample volume (the same electron system) or in separate volumes (different electron systems) due to fluctuation of the dopant concentration across the material. Macroscopic data like resistivity average over the whole sample volume under observation. Local (microscopic) methods see environment of the particular atoms used as probes. The Mössbauer spectroscopy has this property provided resonant atoms are constituents of the material [8-10]. Hence, one can distinguish between homogeneous and inhomogeneous systems within the same crystal phase. The non-magnetic spectral component is due to iron located in the superconducting regions or eventually due to the normal over-doped regions. The magnetic component arises from regions with SDW – non-superconducting. Hence, it seems that superconductivity is restricted to some regions of the samples investigated due to slight variation of the dopant concentration across the bulk of the material [11].

**Table**

Essential results obtained by the Mössbauer spectroscopy at various temperatures $T$ for doped (superconducting) compounds. The square root from the mean squared amplitude of SDW is shown for magnetic components. The symbol $A$ denotes contribution of the non-magnetic (NM) and SDW components. The symbol $S$ stands for the total spectral shift versus room temperature (RT) α-Fe, while the symbol $\Delta$ stands either for the quadrupole splitting (NM components/spectra) or for the first order correction to the hyperfine interaction due to the electric quadrupole interaction (SDW components). The symbol $\Gamma$ denotes absorber line width. Errors are of the order of unity for the last digit shown or they are shown explicitly otherwise.

| $T$ (K) | NM/SDW | $A$ (%) | $S$ (mm/s) | $\Delta$ (mm/s) | $\Gamma$ (mm/s) |
|---|---|---|---|---|---|
| $Ba_{0.7}Rb_{0.3}Fe_2As_2$ | | | | | |
| RT | NM | 100 | 0.41 | 0.07 | 0.20 |
| 145 | NM | 100 | 0.49 | 0.02 | 0.21 |
| 80 | NM | 87 | 0.51 | 0.07 | 0.19 |
|  | SDW 4.4 T | 13(3) | 0.52 | -0.01(2) | 0.15(3) |
| 4.2 | NM | 85 | 0.52 | 0.09 | 0.21 |
|  | SDW 4.9 T | 15(4) | 0.53 | -0.01(2) | 0.17(5) |
| $CaFe_{1.92}Co_{0.08}As_2$ | | | | | |
| RT | NM | 100 | 0.43 | 0.20 | 0.24 |
| 210 | NM | 100 | 0.50 | 0.23 | 0.20 |
| 145 | NM | 56 | 0.55 | 0.25 | 0.16 |
|  | SDW 4.8 T | 44 | 0.49 | -0.10 | 0.15 |
| 100 | NM | 27 | 0.57 | 0.26 | 0.14 |
|  | SDW 5.2 T | 73 | 0.54 | -0.09 | 0.16 |
| 90 | NM | 25 | 0.57 | 0.27 | 0.12 |
|  | SDW 5.3 T | 75 | 0.55 | -0.09 | 0.17 |
| 80 | NM | 23 | 0.58 | 0.28 | 0.11 |
|  | SDW 5.4 T | 77 | 0.55 | -0.09 | 0.18 |
| 4.2 | NM | 17 | 0.58 | 0.27 | 0.16 |
|  | SDW 6.7 T | 83 | 0.56 | -0.04 | 0.16 |

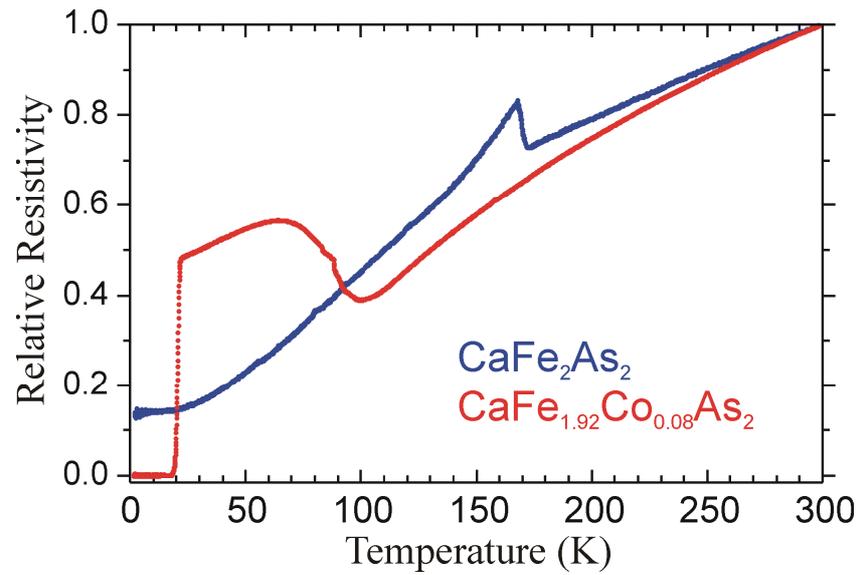

**Figure 1** Resistivity normalized to the resistivity at 300 K (relative) plotted versus temperature for $CaFe_2As_2$ and $CaFe_{1.92}Co_{0.08}As_2$.

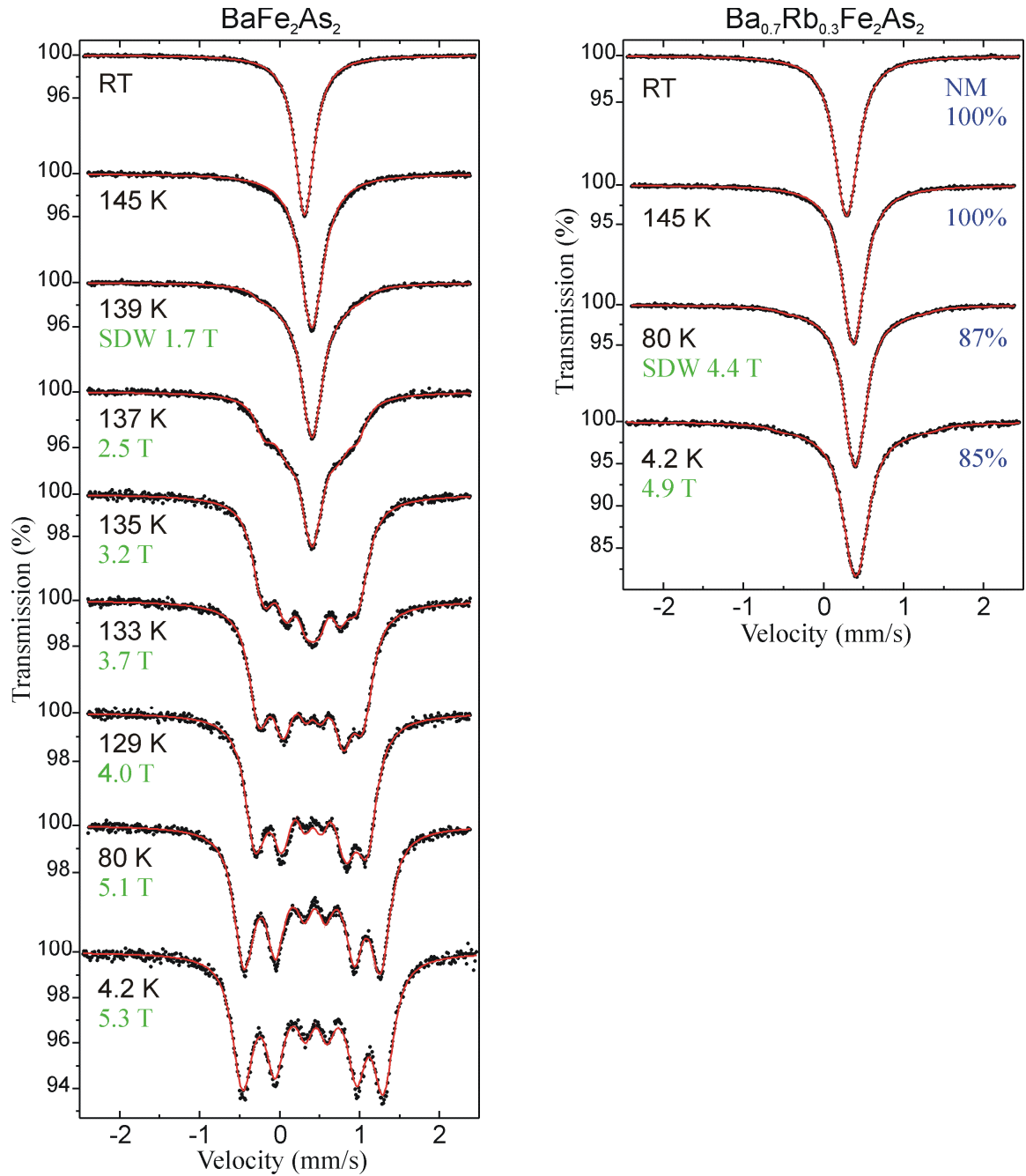

**Figure 2** Mössbauer spectra of the parent compound $BaFe_2As_2$ (a) and doped superconducting compound $Ba_{0.7}Rb_{0.3}Fe_2As_2$ (b) versus temperature. The symbol RT stands for the room temperature. The square root from the mean squared amplitude of SDW is shown in green for spectra with the magnetic component. A contribution of the non-magnetic component (NM) is shown in blue for the doped compound (b).

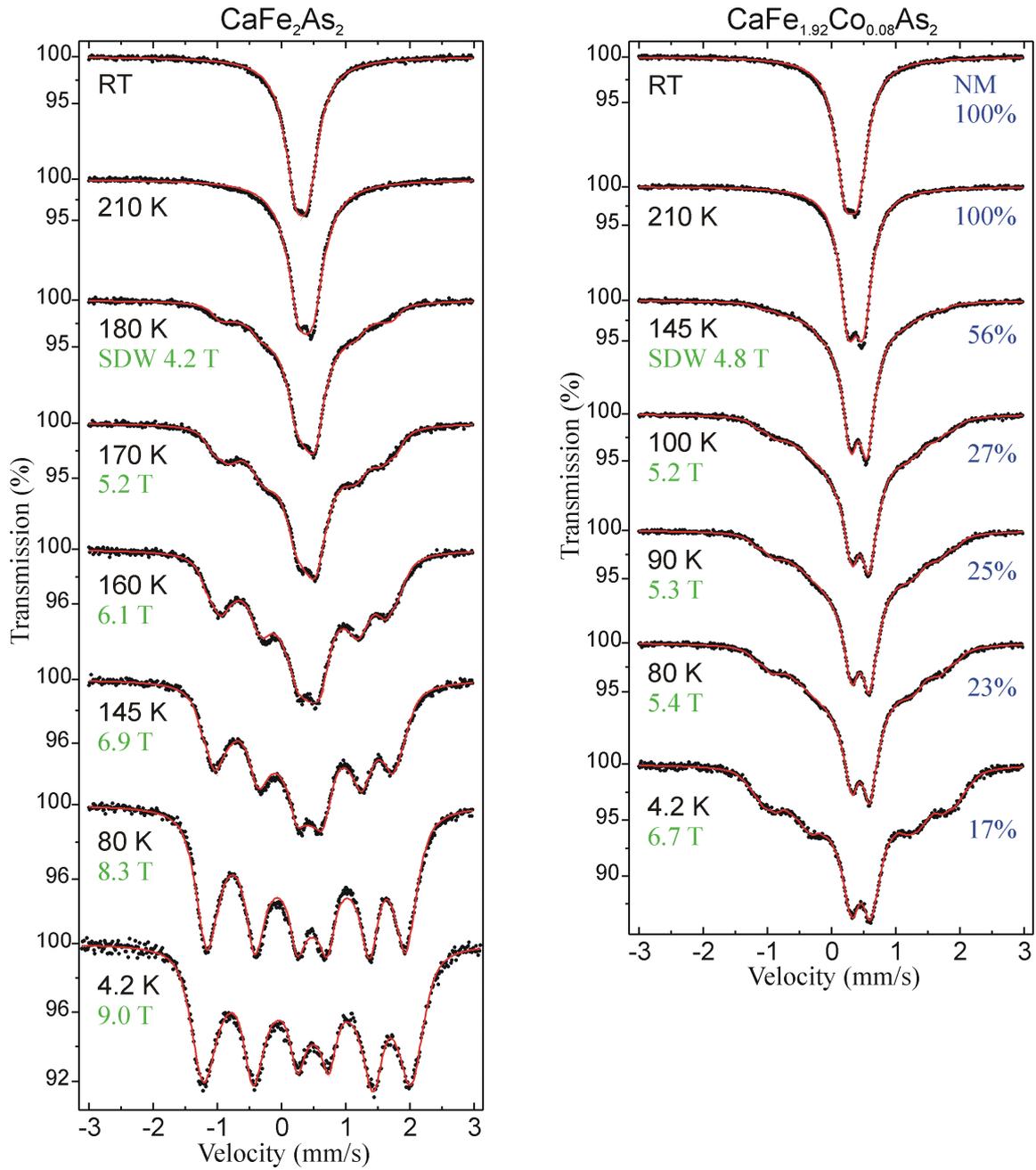

**Figure 3** Mössbauer spectra of the parent compound of $CaFe_2As_2$ (a) and doped superconducting compound $CaFe_{1.92}Co_{0.08}As_2$ (b) versus temperature. The symbol RT stands for the room temperature. The square root from the mean squared amplitude of SDW is shown in green for spectra with the magnetic component. A contribution of the non-magnetic component (NM) is shown in blue for the doped compound (b).

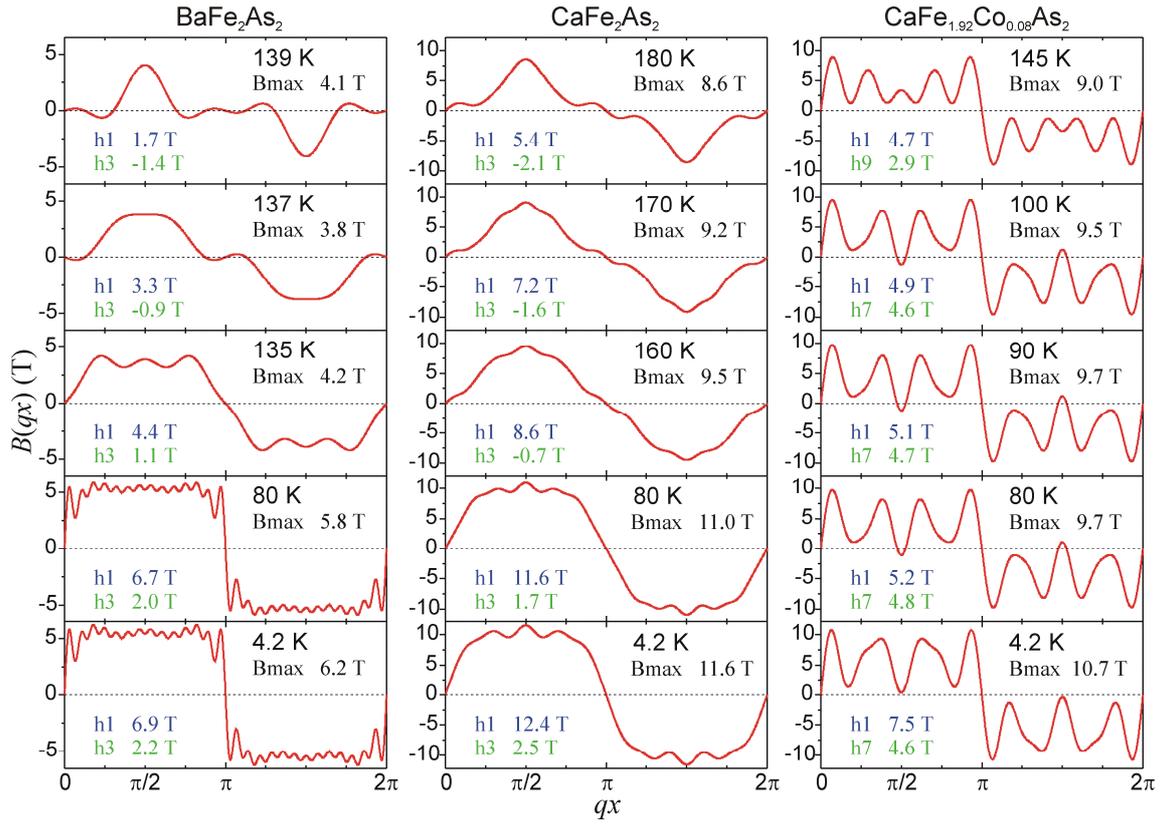

**Figure 4** Shape of SDW plotted for various temperatures for parents and CaFe$_{1.92}$Co$_{0.08}$As$_2$. The symbol B$_{max}$ stands for the maximum SDW amplitude, while the symbols h$_n$ denote amplitudes of two dominant harmonics.

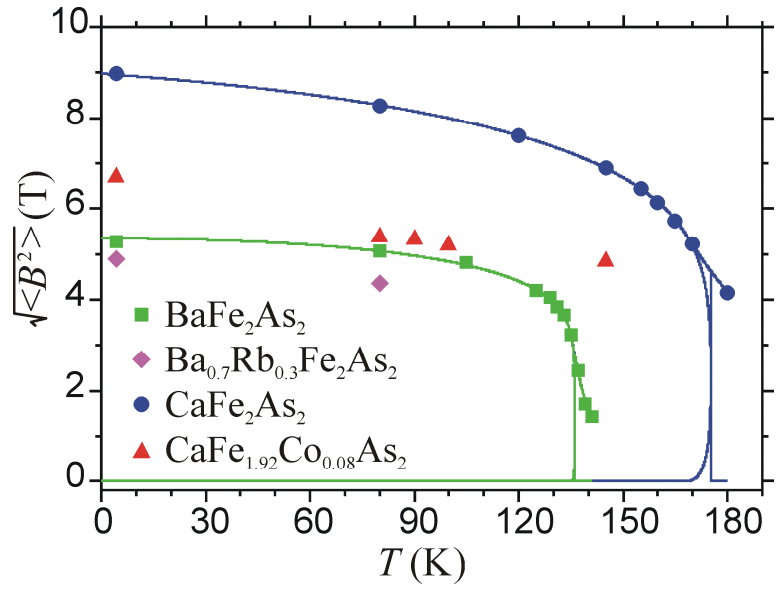

**Figure 5** Square root from the mean squared amplitude of SDW $\sqrt{\langle B^2 \rangle}$ plotted versus temperature. Solid lines represent total, coherent and incoherent contributions for parents [6].